\documentclass[letterpaper,12pt]{article}   
\usepackage{osajnl2} 
\usepackage[draft]{hyperref} 
\usepackage{stmaryrd}
\usepackage{color} 
\usepackage{amsmath}

\begin{document}

\title{Properties of a Variable-delay Polarization Modulator}


\author{David T. Chuss,$^{1,^*}$ Edward J. Wollack,$^{1}$ Ross Henry,$^1$ Howard Hui,$^5$ Aaron J. Juarez,$^{4}$ Megan Krejny,$^{3}$
	S. Harvey Moseley,$^1$ Giles Novak$^{2}$}
\address{$^1$Observational Cosmology Laboratory, Code 665, NASA Goddard Space Flight Center, Greenbelt, MD 20771, USA}
\address{$^2$Northwestern University, Department of Physics and Astronomy, 2131 Tech Drive, Evanston, IL, 60208, USA}
\address{$^3$The Department of Astronomy, University of Minnesota, Minneapolis, MN 55455, USA}
\address{$^4$The University of Texas at Austin, Astronomy Department, Austin, TX 78712, USA}
\address{$^5$ The California Institute of Technology, Pasadena, CA 91125, USA}
\address{$^*$Corresponding author: David.T.Chuss@nasa.gov}

\begin{abstract}
We investigate the polarization modulation properties of a variable-delay polarization modulator (VPM). 
The VPM modulates polarization via a
variable separation between a polarizing grid and a parallel mirror. We find that in the limit where the wavelength is much larger than the diameter of the metal wires that comprise the grid, the phase delay derived from the geometric separation between the mirror and the grid is sufficient to characterize the device. However, outside of this range, additional parameters describing the polarizing grid geometry must be included to fully characterize the modulator response.  In this paper, we report test results of a VPM at wavelengths of 
350  $\mu$m and 3 mm. Electromagnetic simulations of wire grid polarizers were performed and are summarized using a simple circuit model that incorporates the loss and polarization properties of the device.  
\end{abstract}


\maketitle 

\section{Introduction}
Astronomical polarimetry in the far-infrared through microwave portion of the electromagnetic spectrum is a useful tool for probing the physics of interstellar dust, investigating the role of magnetic fields in star formation, and characterizing the radiation from the early universe.  In each of these cases, polarization modulation is an important element of instrument design because it enables a precision measurement  by encoding the polarization information and thus separating it from the typically larger unpolarized background signal.

Variable-delay polarization modulators (VPMs) operate through the introduction of a controlled, variable phase delay between two orthogonal linear polarizations. This is accomplished in recent work by placing a polarizing grid in front of a mirror and varying the separation \cite{Chuss06,Krejny08}, although other related architectures have been used for polarization modulation via introduction of a variable electrical delay \cite{Martin74, Catalano04}.  
For a single frequency, the polarization transfer function of the VPM is a sinusoidal function of the phase delay between the two orthogonal polarizations.  
The key to understanding the polarization transfer function is to determine how the phase delay is related to the grid-mirror separation, since the latter is the quantity that is directly measurable in an instrumental setup. In practice, the polarization response of a specific modulator can be measured to produce a template for use in reducing astrophysical data. As such, the work presented here is not required for
demodulating a time-ordered data set. The utility of the model described here is that it allows calculation of the achievable optical response. It provides both guidance regarding the modulators sensitivity to design parameters and validation metrics for the observed instrumental performance. In addition to VPMs, grid-mirror systems have been implemented in a variety of other applications \cite{Erickson87, Erickson78, Harvey93, Houde01, Akeson96, Shinnaga99, Siringo04, Siringo10, Leonardi06}.  Elements of the analysis presented here are applicable to these systems as well.

In this paper, the polarization transfer function for a single modulator from the Hertz/VPM instrument is measured at 350 $\mu$m and 3 mm. The observed response is well-characterized by a transmission line model.  This work represents a refinement of previous results \cite{Krejny08} in which it was qualitatively shown that VPMs modulate polarization with reasonable 
efficiency and without introduction of major artifacts.  Here, we present new lab tests of Hertz/VPM
collected after improving the optical alignment. We demonstrate 
quantitative agreement between the measured polarization transfer 
function and a transmission line model.  Also, we show 
that the overall modulation efficiency of Hertz-VPM is within $\sim$2\% 
of the expected value at $350$ $\mu$m and in closer agreement at 3 mm.  The organization of the paper is as follows: In section 2, we briefly review the principle of operation of the VPM. In section 3 we describe the circuit model for the VPMs.  In section 4, we discuss the current state-of-the-art for analytical models used to calculate the circuit parameters and include numerical simulations to overcome their limitations. In sections 5 and 6, this model is applied to measurements of the VPM, and the results are discussed in section 7.

\section{The Variable-delay Polarization Modulator (VPM)}
\begin{figure}[htpb]
	\centering
	\includegraphics[width=2in]{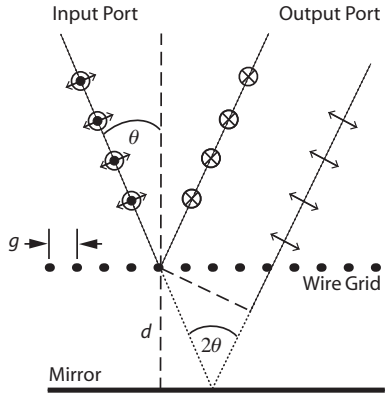}
	\caption{The VPM consists of a polarizing grid placed in front of and parallel to a planar mirror. The polarization parallel to the 
	grid wires is reflected by the grid. The orthogonal linear polarization passes through the grid and is reflected off the mirror. The two 
	components are recombined at the output port with a relative phase delay that is dependent upon the grid-mirror separation, $d$. The wire grid spacing or pitch
	is indicated by $g$.}
	\label{fig:setup}
\end{figure}

The VPM consists of a polarizing grid positioned parallel to and in front of a mirror. A change in grid-mirror separation corresponds to a change in introduced phase between two linear orthogonal polarizations.  The device is show in Figure~\ref{fig:setup}.  Choosing coordinates such that Stokes $Q$ gives the difference between the polarization states parallel and perpendicular to the VPM wires, the polarization transfer function can be expressed as
\begin{equation}
	U^\prime=U\cos{\delta}+V\sin{\delta}.
	\label{eq:stokes}
\end{equation}
Here $U$ and $U^\prime$ are the input and output Stokes $U$ parameter, and $V$ is the input circular polarization. Here $\delta$ is the electrical phase delay between the polarized component transmitted by and that reflected by the grid upon recombination at the output port of the device. This is the phase of interest when using the VPM as a modulator. 

In the limit in which the wavelength is much larger than the length scales that characterize the local grid geometry, the VPM phase is proportional to the path difference between the two polarizations, 
\begin{equation}
	\delta_{\infty}=\frac{4\pi d}{\lambda}\cos{\theta},
\label{eq:geom}
\end{equation}
where $\theta$ is the incident angle, as shown in Figure~\ref{fig:setup}. Here, the $\infty$ subscript indicates that this relationship holds in the long wavelength limit. In many polarizing grid applications the size of the wires relative to the wavelength is large enough to produce measurable deviations from the phase relation given in Equation~\ref{eq:geom}.

We assume throughout this paper that the grid wires are uniform, infinite in extent, and lie in a plane. Diffraction grating lobes are negligible when the wire radius, $a$, and grid constant or center-to-center wire pitch, $g$, are much smaller than the wavelength of the incident radiation.  In this limit, the structure has a homogeneous response for each polarization. 

\section{A VPM Model}

At the highest level, our goal is to describe the polarization modulation properties of the VPM. To do this, several details are worth briefly considering before we proceed.  Both Jones matrices \cite{Jones41, Brosseau} and Mueller matrices allow one to propagate the polarization properties through a system having no reflections between elements.  These two descriptions can be shown to be mathematically equivalent \cite{Sternberg}. A polarization modulator, such as the VPM, that inherently relies on the interference of the fields in the device is not directly amenable to this analysis approach. To incorporate the influence of multiple reflections in such a system,  the ``$ABCD$'' matrix approach can be employed. The $ABCD$ matrix is also known as a transfer, transmission line, chain, or characteristic matrix \cite{Pozar, Goldsmith, Yeh, Palik} in the microwave and optical literature. Physically, this matrix formulation links the propagation of the input and output electric and magnetic fields through a given component.  Once the modulator is modeled with these tools, it is straightforward to calculate a Jones matrix for the entire system \cite{Adachi60}, assuming that the input and output ports are matched.

From an instrumenation perspective, an ``ideal'' polarizing grid perfectly separates orthogonal linear polarization components with a response that is independent of frequency.  That is, the grid perfectly transmits the polarization component having the electric field perpendicular to the grid wires and perfectly reflects the component having the electric field parallel to the grid wires.  The former corresponds to zero impedance contrast between free space and the grid; the latter corresponds to an infinite impedance contrast.  For physically realizable grids,  the achievable impedance contrast for each of the two polarizations depends on the electromagnetic properties and geometry of the structure. In the limit the wire radius and separation are finite compared to the wavelength of the incident radiation, the response becomes dependent on frequency. Motivated by the observation that the wire grid is a polarization- and frequency-selective surface, we model the structure using a transmission line approach. 
 
\begin{figure}[htpb]
	\centering
	\includegraphics[width=5in]{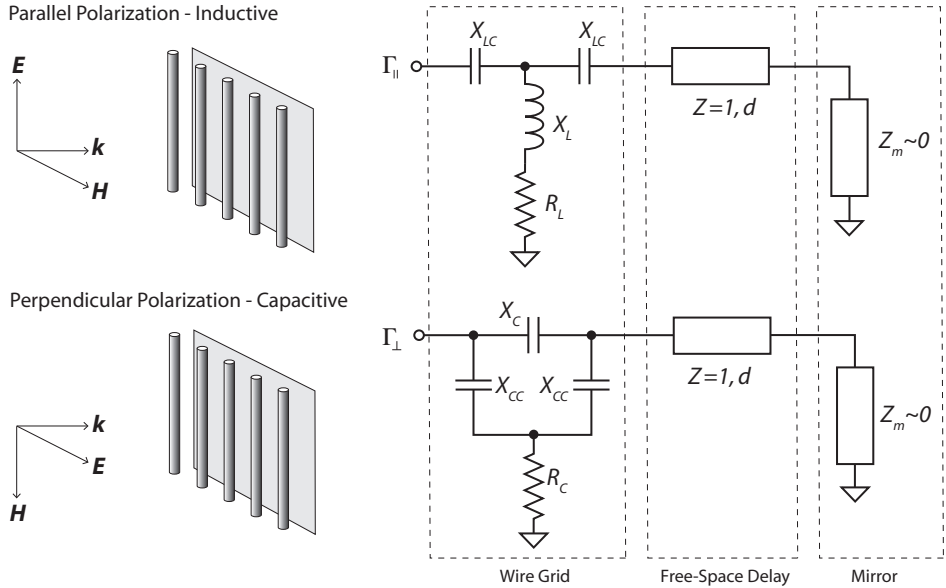}
	\caption{In the limit the wavelength is large compared to the wire pitch, the VPM can be modeled by two independent circuits.  An inductive circuit is used for the polarization component having the electric field parallel to the grid wires (top). A capacitive circuit is used to model the polarization component having the electric field perpendicular to the grid wires (bottom). }
	\label{fig:xmissline}
\end{figure}

We model the polarizing grid using a circuit representation \cite{Marcuvitz} of the VPM as shown in Figure~\ref{fig:xmissline}.  We normalize all impedances to that of free space, $Z_0= 377\,\Omega$ per square.  For the polarizing grid, the transmission and reflection of the electric field component parallel to the grid wires can be represented by an inductive circuit.  The circuit reactances used are defined in Figure~\ref{fig:xmissline}. We compute the $ABCD$ matrix from these as follows:
\begin{equation}
\left(  \begin{array}{cc} 
      1+ \frac{iX_{LC}}{iX_L+R_L} &  \left(2+\frac{iX_{LC}}{iX_L+R_L}\right)iX_{LC}\\
     \frac{1}{iX_L+R_L}& 1+ \frac{iX_{LC}}{iX_L+R_L} \\
   \end{array}\right).
\end{equation}
Likewise, the grid's effect on the polarization component having its electric field perpendicular to the grid wires can be represented by a capacitive circuit and is
given by 
\begin{equation}
\frac{1}{iX_{CC}+\left(2+\frac{X_C}{X_{CC}}\right)R_C}
\left(  \begin{array}{cc} 
      iX_{C}+iX_{CC}+\left(2+\frac{X_C}{X_{CC}}\right)R_C& (2R_C+iX_{CC})iX_C\\
     2+ \frac{X_C}{X_{CC}}& \left(2+\frac{X_C}{X_{CC}}\right)R_C+\left(1+\frac{X_C}{X_{CC}}\right)iX_{CC} \\
   \end{array}\right).
\end{equation}

\begin{figure}[htpb]
	\centering
	\includegraphics[width=3.5in]{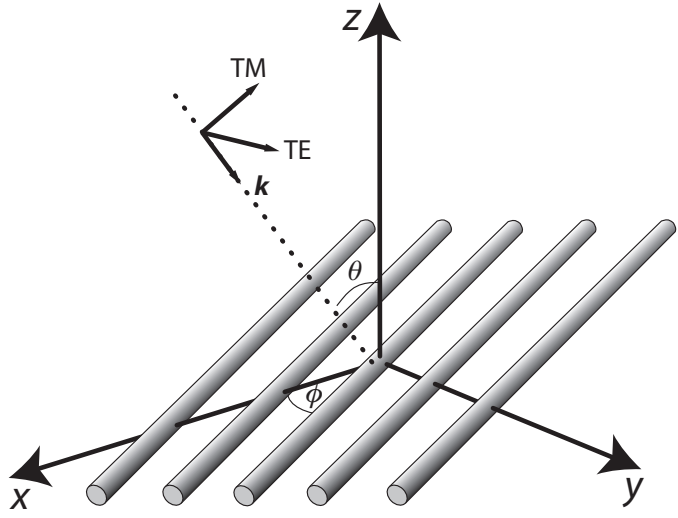}
	\caption{The geometry for the general polarizing grid lying in the x-y plane is shown.}
	\label{fig:coordsys}
\end{figure}

Each of these matrices can be cascaded with the transmission matrices for 
a free space delay and a short.  The resulting matrix is employed to calculate the reflection coefficients for each polarization,  $\Gamma_{\parallel}$ and $\Gamma_{\perp}$, as a function of frequency.  The Jones matrix for the VPM can then be expressed as 
\begin{equation}
\overline{J}=\left(  \begin{array}{cc} 
      \Gamma_{\parallel}& 0\\
      0 & \Gamma_{\perp}
      \end{array}\right).
\end{equation}
The density matrix describing the polarization state may be written as a linear combination of Stokes parameters ($I$, $Q$, $U$, $V$) {\cite{Brosseau}.
\begin{equation}
D=I\left(\begin{array}{cc}1& 0\\0& 1\end{array}\right)+
Q\left(\begin{array}{cc}1& 0\\0& -1\end{array}\right)+
U\left(\begin{array}{cc}0& 1\\1& 0\end{array}\right)+
V\left(\begin{array}{cc}0& -i\\i& 0\end{array}\right).
\end{equation}
The density matrix of the incoming light can be mapped to that of the outgoing light, $D^\prime$, by using the system's Jones matrix
$\overline{D}^\prime=\overline{J}^\dagger\overline{D}\,\overline{J}.$
This model is applicable for normal incidence. In the more general case of non-normal incidence, it is convenient to parameterize the model in terms of the incidence angle, $\theta$, and the angle between the grid wires and the plane of incidence, $\phi$. See Figure~\ref{fig:coordsys}.
This general case can be calculated using the above formalism along with a procedure similar to that described by Goldsmith \cite{Goldsmith} where the free-space TE and TM polarizations are projected onto the grid wires.  The polarization separation then occurs in the basis of the wires, as done above. 

If the incidence angle is non zero, there are two effects. First, the length of the grid-mirror separation in the transmission line is multiplied by $\cos{\theta}$.  Second, it is necessary to take into account the differences at the grid interface for the TM and TE polarizations.  For the case of the transverse magnetic (TM) polarization mode, the effective impedance is altered from the free-space value by a factor of $\cos{\theta}$.  This is due to the fact that the electric field that is projected onto the plane is smaller than the freely-propagating value.  For the transverse electric (TE) polarization, it is the magnetic field vector that is projected into the plane and so the impedance differs from that of free space by a factor of $1/\cos{\theta}$.  

A simple case that one can consider for non-zero incidence angle is that where the wires are either parallel or perpendicular to the plane of incidence. In this case, each of the TE and TM modes correspond to only one of the circuits described above. For example, if the wires are parallel to the plane of incidence, the TE and TM modes correspond to the capacitive and inductive circuit elements, respectively.
For a general grid rotation angle $\phi$, four circuit models must be considered utilizing all of the combinations of (TM,TE) and (inductive, capacitive).  After the transmission matrices are found for the two transmission lines in Figure~\ref{fig:xmissline}, scattering parameters can be determined. Once the scattering matrices are found for the VPM in each of these four cases, we can combine the results into a  single Jones matrix,
\begin{align}
\nonumber \overline{J}(d)&=
\left(  \begin{array}{cc} \cos\phi & -\sin\phi\\ \sin\phi & \cos\phi \end{array}\right)
\left(  \begin{array}{cc} 1 & 0\\ 0 & -1 \end{array}\right)
\left(  \begin{array}{cc}  \Gamma_{\parallel}^{TM}(\theta,d)\cos\phi &  \Gamma_{\parallel}^{TE}(\theta,d)\sin\phi \\ -\Gamma_{\perp}^{TM}(\theta,d)\sin\phi 
& \Gamma_{\perp}^{TE}(\theta,\phi)\cos\phi \end{array}\right)
\\
&=\left(  \begin{array}{cc} 
      \Gamma_{\parallel}^{TM}(\theta,d)\cos^2{\phi}-  \Gamma_{\perp}^{TM}(\theta,d)\sin^2{\phi}& \left({\Gamma_{\parallel}^{TE}(\theta,d)+\Gamma_{\perp}^{TE}(\theta,d)}\right)\cos{\phi}\sin{\phi}\\
      \left({\Gamma_{\parallel}^{TM}(\theta,d)+\Gamma_{\perp}^{TM}(\theta,d)}\right)\cos{\phi}\sin{\phi} & \Gamma_{\parallel}^{TE}(\theta,d)\sin^2{\phi}-  \Gamma_{\perp}^{TE}(\theta,d)\cos^2{\phi}
   \end{array}\right).
\end{align}
This Jones matrix maps the incident electric field $\left({\begin{array}{c c} E^{TM} & E^{TE}\end{array}}\right)^T$ to the reflected field  $\left({\begin{array}{c c} E^{TM\prime} & E^{TE\prime}\end{array}}\right)^T$. The resultant polarization transformation for the system can be computed from
$\overline{D}^\prime=\overline{J}^{\dagger}\overline{D}\,\overline{J}$.

\section{Circuit Parameter Values \label{sec:values}}

The circuit topology described above is physically motivated; however, circuit parameters for the grid are required to analyze the response of the VPM.  Many treatments found in the literature for wire grid polarizers consider a simple analytically-treatable limiting case \cite{Marcuvitz, Wait54, Larsen62, Houde01}.  These models assume no diffraction, $\lambda>2g$, little or no azimuthal dependence of current density on the wires, $a\ll\lambda$, and that the grid filling fraction is small, $2a/g< 1/2\pi$. 
However, at millimeter and submillimeter wavelengths, realizable grid structures often violate the third condition as indicated in Figure~\ref{fig:Marc}. A survey of practical grid polarizers indicates that filing fractions in the range $0.3 < 2a/g < 0.8$ are available and achieve high polarization isolation. Such grids are outside of the range of applicability of the perturbation-based models referred to above. An interesting extension of this approach utilizing higher-order impedance boundary conditions applicable for sparse grids has been explored \cite{Yatsenko00}. Green's function and lattice-sum \cite{Chambers80, Yasumoto99, Kushta00} methods enable more accurate treatments without a need for these approximations. Here we employ a hybrid numerical approach that allows us to analyze the behavior of wire grid polarizer used for this work.  In this section, we first revisit the classic analytical treatment of polarizing grids and motivate the need for a more rigorous treatment.

\subsection{Sparse Grid Approximation}

As an example of commonly-employed analytical methods, we explicitly give the expressions for the circuit elements from Marcuvitz \cite{Marcuvitz}; however, this is representative of the approximations used in quasioptical treatments of grids.  In the following, all reactances are normalized with respect to the impedance of free space.  Given a wire radius $a$ and a separation or ``pitch'', $g$, the 
circuit elements for the polarization parallel to the wires (the ``inductive'' mode; Marcuvitz \S 5.21, ``Inductive Posts'') are given by

\begin{align} 
X_{LC}&=-  \frac{g}{\lambda}\left(\frac{2\pi a}{g}\right)^2\\
X_L&= \frac{g}{\lambda}\left\{\ln\frac{g}{2\pi a}+\sum_{m=1}^{\infty} \left(\frac{1}{\sqrt{m^2-\left(\frac{g}{\lambda}\right)^2} }-\frac{1}{m}\right)\right\}.
\label{eq:indmode}
\end{align}
The series for the inductive reactance in Equation~\ref{eq:indmode} is rapidly convergent. In the limit of applicability, this series can be approximated by $\sim0.6\cdot (g/\lambda)^2$ and is a small perturbation. As a result, the factor in curly braces, which we identify with the quasi-static inductance, is essentially frequency-independent.

For the polarization having the electric field perpendicular to the wires (the ``capacitive'' mode; Marcuvitz \S 5.13, ``Capacitive Posts''), the reactive circuit elements are given for a symmetrically-centered capacitive post \cite{Marcuvitz}:
\begin{align}
X_{CC}&=-\frac{\lambda}{g}\left\{2\left(\frac{g}{2\pi a}\right)^2 A_2\right\}\\
X_{C}&=-\frac{\lambda}{g}\left\{2\left(\frac{\lambda}{g}\right)^2\left(\frac{g}{2\pi a}\right)^2 A_1-\frac{1}{4}\left(\frac{2\pi a}{g}\right)^2\frac{1}{A_2}\right\}^{-1},
\end{align}
where $A_1$ and $A_2$ for normal incidence are defined as
\begin{align}
A_1=1+&\frac{1}{2}\left(\frac{2\pi a}{\lambda}\right)^2\left\{\ln\left(\frac{g}{2\pi a}\right) +\frac{3}{4}+
\sum_{m=1}^\infty \left(\frac{1}{\sqrt{m^2-\left(\frac{g}{\lambda}\right)^2}}-\frac{1}{m}\right)\right\}\\
\nonumber A_2=1-&\frac{1}{2}\left(\frac{2\pi a}{\lambda}\right)^2\left\{\ln\left(\frac{g}{2\pi a}\right)-\frac{11}{4}\right\}
\\ 
& +\left(\frac{2\pi a}{g}\right)^2\left\{\frac{1}{24}-\sum_{m=1}^\infty \left(m-\frac{1}{2m}\left(\frac{g}{\lambda}\right)^2
-\sqrt{m^2-\left(\frac{g}{\lambda}\right)^2}\right)\right\}.
\end{align}

The circuit parameters given above are plotted as functions of the geometric filling fraction, $2a/g$, in Figure~\ref{fig:Marc}.  In plotting the circuit parameters, we multiply the reactances by a factor of $\lambda/g$ or $g/\lambda$ for inductive and capacitive components, respectively.  This separation between wavelength-dependence and the structure's geometry is anticipated in the quasistatic limit (i.e. the limit in which each circuit element in the model can be treated as a discrete frequency-independent component). The triangles at the bottom of the figure indicate typical values of grids employed in the submillimeter and millimeter, and the vertical gray line marks the position at which $2a/g=1/\pi$. We note that the expression for the shunt inductance is zero at this point and negative beyond it. This is a manifestly unphysical result, as the wires do not actually become ``capacitive'' for the parallel polarization, but approach the inductance of a corrugated conductive plate. Thus, a different approach is required in this regime, as the limit of applicability indicated by Marcuvitz is violated.

  \begin{figure}[htpb]
	\centering
	\includegraphics[width=4in]{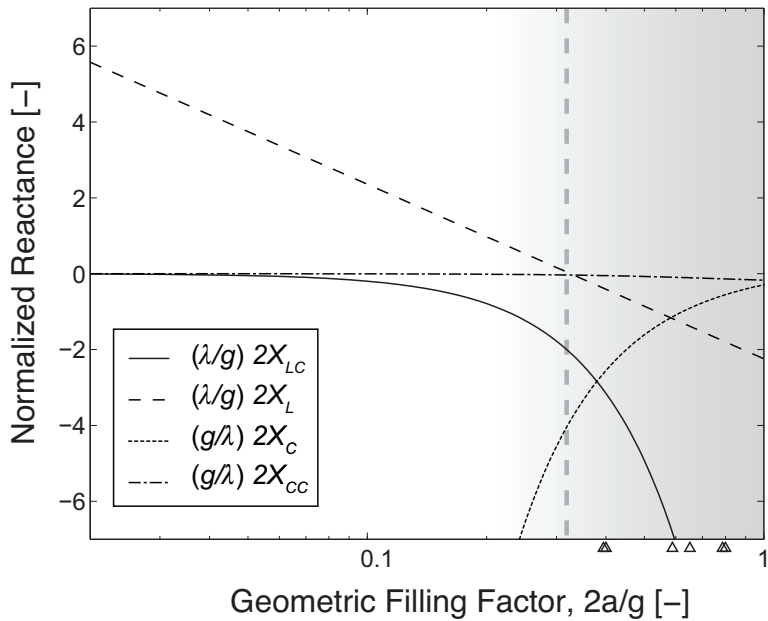}
	\caption{The Marcuvitz circuit parameters are plotted as functions of the geometric filling fraction, $2a/g$.  The vertical gray line is shown at $2a/g=1/\pi$. The shaded region indicates the region in which the Marcuvitz approximation ceases to hold \cite{Marcuvitz}. Triangles indicate typical values of $2a/g$ for grids constructed for submillimeter and millimeter use. }
	\label{fig:Marc}
\end{figure}

\subsection{Numerical Simulations of Grid Performance}

In order to understand the inductance and loss of wire grids, we perform numerical simulations as a function of filling fraction over the range $0.02<2a/g<1$.  The numerical grid simulations were carried out with CST (Computer Simulation Technology) MICROSTRIPES$^\mathrm{TM}$ Time-Domain Transmission-Line Matrix (TLM) solver and Ansoft HFSS (High Frequency Structural Simulator). See Figure~\ref{fig:unitcell} for the electromagnetic configuration and the boundary conditions used for modeling each polarization. We consider the grid's absorption from the structure in the limit where the field penetration depth is small compared to the wire diameter and the incident wavelength is greater than two times the wire pitch. In this limit the higher order Floquet harmonics or grating lobe responses are absent, and the fields are quasi-static in nature and a perturbation on the lossless case.
 
Wheeler's incremental inductance rule \cite{Edwards} can be applied to compute the resistive losses from the wires in the grid. This approach implicitly assumes that the current distribution on the conductors does not vary appreciably over distances comparable to the thickness of the wire and is a convenient parameterization.  This condition is satisfied for a wire having a well-developed skin effect, $\delta \ll a$, in the single mode limit. The anticipated form can be expressed as a product of two separable functions: the first dependent on the grid filling factor, $2a/g$, and the second a function of skin depth over the wavelength, $\delta/\lambda$. Guided by this physical insight, we proceed to model the grid in the single mode approximation to further explore the scaling properties of the grid inductance and loss.
 
  \begin{figure}[htpb]
	\centering
	\includegraphics[width=6in]{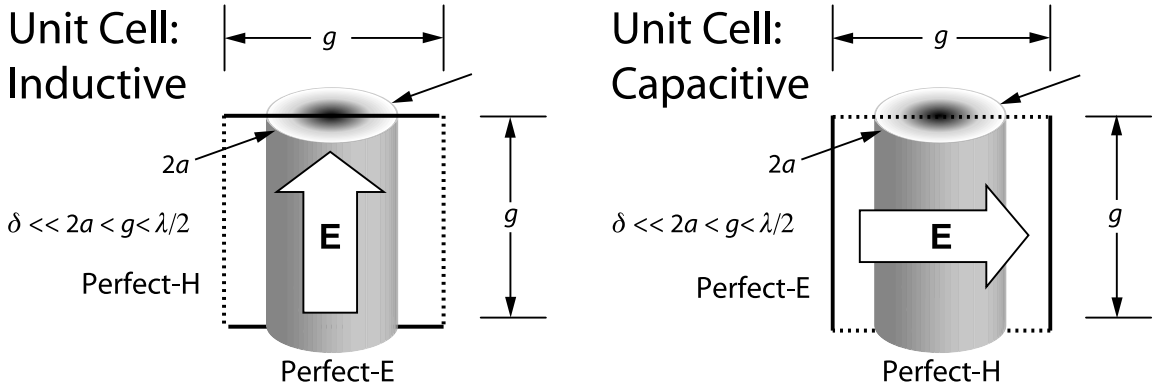}
	\caption{The unit cells used for the simulation in both the inductive (left) and capacitive (right) modes are shown. The large arrow in each case indicates the direction of the incident electric field in the simulation.  Perfect-``E'' and -``H'' indicate the use perfect electric and magnetic mirrors on the boundaries of the unit cell, respectively.  Plane wave illumination of the unit cell with these boundary conditions allows representation of an infinite grid for $\lambda>2g$. }
	\label{fig:unitcell}
\end{figure}

The power reflection coefficient, $R$, from the impedance presented by the shunt reactance of the grid, $Z_L=R_L+iX_L$, is: 
\begin{equation}
R=|S_{11}|^2=\left|\frac{-1}{1+2Z_L}\right|=\frac{1}{(1+2R_L)^2+(2X_L)^2},
\end{equation}
where all impedances are normalized to $Z_0$. The resulting reflection phase is 
\begin{equation}
\phi_r=\pi-\arctan{\left(\frac{\Im(2Z_L)}{1+\Re(2Z_L)}\right)}=\pi-\arctan{\left(\frac{2X_L}{1+2R_L}\right)}.
\end{equation}
For a near optimal grid geometry, the impedance contrast with respect to free space, $2Z_L\ll1$, and the resultant reflection phase is approximately $\pi$.  Similarly, the power transmission can be expressed as
\begin{equation}
T=|S_{21}|^2=\left|\frac{2Z_L}{1+2Z_L}\right|^2=\frac{(2R_L)^2+(2X_L)^2}{(1+2R_L)^2+(2X_L)^2},
\end{equation}
and the power absorption, $A=1-R-T=1-|S_{11}|^2-|S_{21}|^2$, is computed from conservation of energy. Here, $S_{ij}$ are the complex scattering parameters for the structure. A corresponding set of expressions can be written for the capacitive case. We note the ratio of the absorptance over the reflectance can be expressed as
\begin{equation}
\frac{A}{R}=4R_L=\eta\frac{4R_S}{Z_0}=\eta\frac{4\pi\delta}{\lambda},
\end{equation}
where we recall from Figure~\ref{fig:xmissline} that $R_L$ is the inductive mode's resistance normalized to the impedance of free space. The bulk surface resistivity of the metal is $R_S=1/\sigma\delta$, $\eta$ is the grid loss efficiency, and $\lambda$ is the observation wavelength. The field penetration depth, $\delta=\sqrt{2/\mu\sigma\omega}$, is computed from the metal's permeability, $\mu$, electrical conductivity, $\sigma$, and the observation frequency, $\omega$. It is informative to consider the dimensionless ratio that occurs in the circuit model above in terms of this parameterization:
\begin{equation}
\frac{4R_S}{Z_0}=\frac{4\sqrt{\omega\mu/2\sigma}}{\sqrt{\mu/\epsilon}}=\frac{4\pi\delta}{\lambda}.
\label{eq:circuit}
\end{equation}
The expression on the right is related to  the Hagen-Ruben emissivity formula \cite{Stratton}. This can be derived from Fresnel coefficients and identified as the emissivity from a bulk metal at normal incidence. This approximation is valid in the limit of low emissivity.  Equation~\ref{eq:circuit} links the circuit theory to the interaction between the materials and the electromagnetic fields. We see that the loss efficiency, $\eta=(A/R)(4\pi\delta/\lambda)^{-1}$, is the loss of the grid relative to a flat sheet made of the same material. We compute the scattering parameters as a function of grid filling factor and frequency and derive equivalent circuit parameters. In Figure~\ref{fig:loss2} we show the grid loss factor, the grid shunt reactance, and the grid reflection phase appropriately normalized to remove the wavelength dependence as a function of grid filling factor. This scaling is anticipated and observed in the quasi-static limit for these circuit elements.
\begin{figure}[htpb]
	\centering
	\includegraphics[width=4.5in]{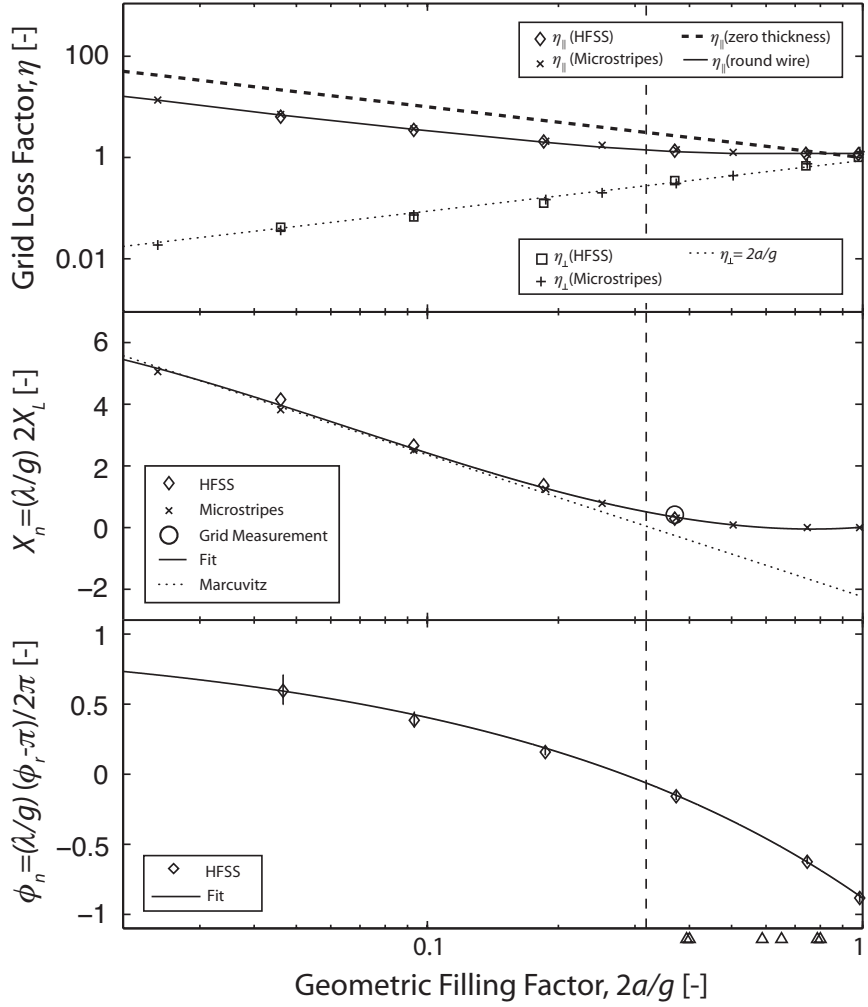}
	\caption{(TOP) The grid loss factor, $\eta$, is plotted as a function of geometric filling factor for a wire grid. Both HFSS and Microstripes simulations are shown for the cases of the electric field perpendicular to and parallel to the grid wires. The filling factors for typical polarizing grids are plotted for comparison. (MIDDLE) The reactance for the inductive mode is shown as a function of the geometric filling factor. A useful interpolation function is  $X_n\cong 0.51-1.19\ln(2\pi a/g)+0.53 (\ln(2\pi a/g))^2+ 0.11(\ln(2\pi a/g))^3$. For comparison, the Marcuvitz model is shown as well as the measured inductance for the grid. (BOTTOM) The normalized reflection phase of the inductive mode is plotted as a function of geometric filling factor for the HFSS simulation. In this case, we find $\phi_n\cong 1-\sqrt{3.545(2a/g)}$.  In each of the panels, the condition $2a/g=1/\pi$, is denoted by a dashed vertical line.  Triangles indicate typically manufactured grid geometries.}
	\label{fig:loss2}
\end{figure}

For the perpendicular illumination case, we find a grid loss factor, $\eta_\perp=2a/g$. For the parallel illumination case, 
some care is required -- the conductive wires effectively define a finite thickness aperture. For this configuration, the influence of the waveguide cutoff has a non-negligible influence on the grid's absorption properties, and we find its influence results in a grid loss factor of $\eta_{\parallel}={\gamma}/{ \tanh(\gamma\pi\cdot 2 a/g)}$ for this polarization. Here, $\gamma\cong1.15$. In referring to Figure~\ref{fig:loss2} we note that this differs from previous grid loss factors reported in the literature. For example, for infinitely thin strips, Ulrich \cite{Ulrich67} finds $\eta_{\parallel}=g/2a$. For small filling factors, this expression and the one derived above have the same slope; however, in the limit $2a/g\rightarrow1$, the influence of the exclusion of the fields by the grid wires has a pronounced influence on the response. This behavior is also observed in our numerical simulations. We note that within the context of the planar waveguide model \cite{Edwards}, by symmetry, the field configuration for this polarization is physically equivalent to that of a microstrip via \cite{Goldfarb91}.

The numerical approach described here was pursued to overcome and understand the limitations of the perturbation expansions in  $a/\lambda$ \cite{Marcuvitz, Wait54, Larsen62} and $\delta/\lambda$ \cite{Houde01}.  The normalized reactance, $X_n$, of the inductive component is computed in the range of interest and is plotted as a function of $2a/g$ in the middle panel of Figure~\ref{fig:loss2}. For the Marcuvitz model that is also shown, $X_n$ becomes negative for $2a/g>1/\pi$ which is a symptom of a failure of the asymptotic expansion and anticipated from the restrictions on the analytical expression as specified \cite{Marcuvitz}. This generic behavior is present in other analytical treatments that are expansions in $a/\lambda$ \cite{Wait54,Larsen62, Houde01} and anticipated since the derived expressions only have logarithmic accuracy. Higher order terms are required in this regime, e.g., see Landau \S33 \cite{Landau}, and \cite{Grover}. However, as $g/\lambda\rightarrow 0$, $|X_L|\rightarrow 0$ as well, and as a result, this particular modeling detail is of reduced importance. 
This analysis suggests that the observed monotonic flattening of $\eta_{||}$ and $X_n$ at larger values of $2a/g$ result in high performance grids; however, the reflectivity and transmissivity are more effectively improved by reductions in $g/\lambda$.


\section{350 $\mu$m VPM Tests}

\subsection{Experimental Setup}
Preliminary characterization of the Hertz VPM instrument has been reported \cite{Krejny08}. Our setup is similar to that used previously and is schematically depicted in Figure~\ref{fig:submm}.  A chopped blackbody is polarized by a wire grid and focused onto the Hertz polarimeter that simultaneously measures horizontal and vertical polarizations.  The VPM under test is placed near a pupil in a collimated beam within the fore optics that couple the source to the Hertz cryostat. For a given grid-mirror separation, the polarization is modulated by rotating a cold half-wave plate (HWP) in discrete steps. The aperture of the blackbody is sized to roughly fill a single Hertz detector. Inside of the Hertz cryostat, the polarization is separated into two linear orthogonal components by a wire grid analyzer. Both the transmitted (T) and reflected (R) components are detected by bolometers.
\begin{figure}[htpb]
	\centering
	\includegraphics[width=4in]{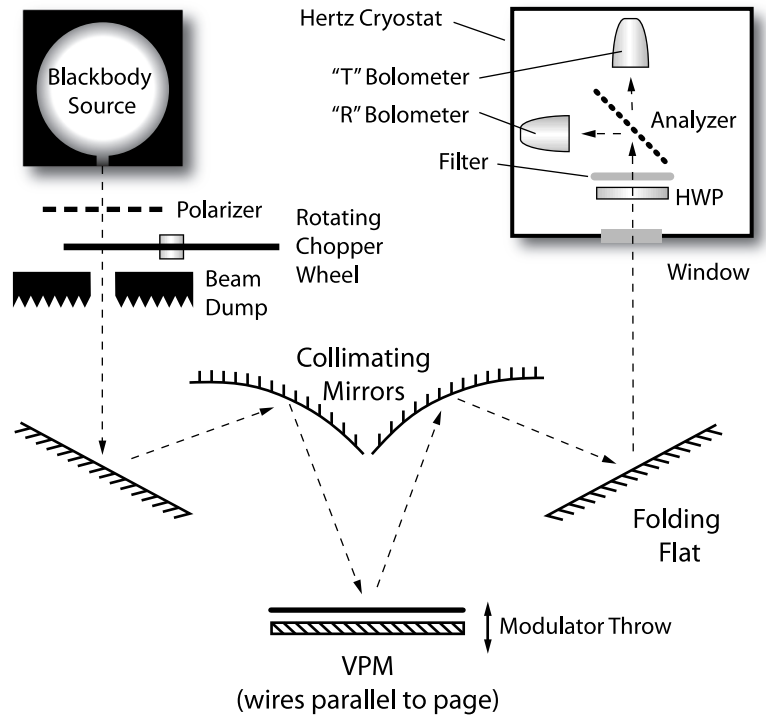}
	\caption{The main elements of the experimental setup for the 350 $\mu$m test are shown \cite{Krejny08}.  Radiation from a blackbody source is polarized by a wire grid polarizer having wires oriented at an angle of 45$^\circ$ with respect to the plane of the page. A chopper modulates the intensity of the signal.   The radiation is collimated prior to being processed by the VPM.  Upon exiting the VPM, the radiation is relayed to the Hertz cryostat. Inside, the radiation passes through the half-wave plate and the bandpass filter before being diplexed into two orthogonal linear polarizations. A bolometer detects the signal in each polarization.}
	\label{fig:submm}
\end{figure}

The polarization transfer function of the VPM is measured as follows. At various grid-mirror separations, the output linear polarization is measured using the Hertz polarimeter. The normalized Stokes parameters are then reported as a function of the grid-mirror separation.  

\subsection{Data Reduction}

At each grid-mirror separation, the basic data analysis pipeline is similar to that described in \cite{Hildebrand00}. However, we describe the process in some detail since there are some deviations that are associated with the difference between laboratory measurements and those done astronomically.  At a single HWP position, 16 ``frames''  are recorded. Each frame consists of two co-added ``chops'' or on-off differences. In our laboratory setup, the chopper phase was observed to drift relative to the commanded TTL signal from the data system. To correct for this, the in-phase data and the quadrature data were both used and the signals for a single frame in the $R$ and $T$ arrays are
\begin{align}
r_i(\alpha_j)&=\left(\left[r_i^{in\,phase}(\alpha_j)\right]^2+\left[r_i^{quad}(\alpha_j)\right]^2\right)^\frac{1}{2}\\
t_i(\alpha_j)&=\left(\left[t_i^{in\,phase}(\alpha_j)\right]^2+\left[t_i^{quad}(\alpha_j)\right]^2\right)^\frac{1}{2}.
\end{align}
Here, $\alpha_j$ is the value of the HWP angle at the $j$th HWP position.
This technique mitigates the problems associated with phase drift in the chopper.

The frames recorded for each HWP are averaged separately for each of the $R$ and $T$ signals,
\begin{align}
 R(\alpha_j)&=\frac{1}{N}\sum_{i=1}^N{r_i}(\alpha_j)\\
T(\alpha_j)&=\frac{1}{N}\sum_{i=1}^N{t_i}(\alpha_j).
\end{align}
Here, $N=16$ is the number of data frames taken at each HWP position. This process is repeated for each of six HWP positions (nominally, 0$^\circ$, 30$^\circ$, 60$^\circ$, 90$^\circ$, 120$^\circ$, and 150$^\circ$). The relative gain between the $R$ and $T$ bolometers is calculated as the ratio of the sum of all the data in the $R$ bolometer and that of the $T$ bolometer:
\begin{equation}
f=\frac{\sum_jR(\alpha_j)}{\sum_jT(\alpha_j)}.
\end{equation}
The polarization signal is then calculated using both the $R$ and $T$ bolometers. This combination aids in the removal of common-mode noise in the system,
\begin{equation}
S(\alpha_j)= \frac{R(\alpha_j)-fT(\alpha_j)}{R(\alpha_j)+fT(\alpha_j)}.
\label{eq:polsig}
\end{equation}

For each grid-mirror separation, the HWP polarization signal in Equation~\ref{eq:polsig} was fit to the modulation function
\begin{align}
S(\alpha)=&\epsilon_{1}(q\cos{4\alpha}+u\sin{4\alpha})+\epsilon_2\cos{2\alpha}+\epsilon_3\sin{2\alpha}.
\label{eq:fit}
\end{align}
The parameters $q\equiv Q/I$ and $u\equiv U/I$ are the normalized linear Stokes parameters in a rotational basis for which $q$ defines the difference in polarization components perpendicular to and parallel to the grid wires of the VPM. We have explicitly included the linear polarization efficiency, $\epsilon_1$.  The system was found to be dominated by drifts on the time scale associated with the HWP cycle.  To account for this, error estimates are determined by goodness-of-fit of the data to Equation~\ref{eq:fit}. 

For the purposes of fitting, $\epsilon_2$ and $\epsilon_3$, are orthogonal to those of importance, $q$ and $u$ and thus are discarded for the primary analysis. However, to verify our understanding of the system, it is worth briefly exploring their physical origin that stems from well-understood non-ideal behavior of the HWP \cite{Savini09}. 

Because the zero angle of $\alpha$ is set by the VPM grid wires and not the fast axis of the HWP, $\epsilon_2$ and $\epsilon_3$ are linked by a rotation to coupling to both unpolarized flux and circular polarization,
\begin{align}
\left(\begin{array}{c}\epsilon_2\\ \epsilon_3 \end{array}\right)=
\left(\begin{array}{cc}\cos{2\chi} & \sin{2\chi}\\ -\sin{2\chi}& \cos{2\chi}\end{array} \right)
\left(\begin{array}{c}\epsilon_2^\prime \\ \epsilon_3^\prime \end{array}\right).
\end{align}
Here, $\chi$ is the angle between the VPM wires and the fast axis of the HWP. The factor $\epsilon_2^\prime$ is the coupling between unpolarized flux and the polarization signal due to bi-attenuance in the HWP. The factor $\epsilon_3^\prime$ is proportional to the circular polarization (Stokes $V$) incident on the HWP. The response of a single-layer HWP, being a function of  the birefringence of the crystal ($\Delta n$) and its thickness, is only ideal at a single frequency.  Deviation from perfect matching between the HWP center frequency and that of the band-defining filter as well as the details of the input spectrum can both cause some leakage between circular and linear polarizations.  

The VPM provides a means for separating the two effects. As the grid-mirror separation is varied, $\epsilon_3^\prime$ will vary with phase delay while $\epsilon_2^\prime$ will be unmodulated.  An unmodulated offset was not observed, implying an upper limit of $\epsilon_2^\prime<0.01$.  From the modulated signal, we find $\epsilon_3^\prime\sim0.2$. At this level, a corresponding reduction of the HWP response to linear polarization is expected.  We calculate this effect to be $\sqrt{1-\epsilon_3^{\prime 2}}=0.98$, thus resulting in an efficiency reduction of $\sim2\%$. This is consistent with previous characterization of the Hertz polarimeter \cite{Schleuning97, Dowell98}.  In the test configuration with the VPM, the HWP was exposed to much larger amounts of circular polarization than in an astronomical instrumentation setting, thus emphasizing this otherwise small effect. 

\subsection{Systematic Offset Removal}

In our test configuration, there are three aluminum mirrors between the source grid and the VPM having incidence angles $\sim20^\circ$.  From this we estimate an induced polarization of  $<0.3\%$ that is primarily in the $\pm q$ direction. Instrumental polarization between the VPM and half-wave plate would lead to a small change in the efficiency of the system. From previous laboratory measurements \cite{Krejny08} we estimate a resulting systematic uncertainty in the polarization efficiency of $<1\%$.

Initially we observed that Stokes $q$ exhibited a modulation as a function of grid-mirror separation that had a similar modulation pattern to that anticipated for circular polarization.  This effect was observed at the 5\% level. By rotating the Hertz cryostat by 90$^\circ$ and remeasuring, this effect was found to switch signs, indicating some birefringence in either the lenses or the pressure window in the cryostat.  Since the axis of such a birefringent element could a priori be oriented in any orientation, a similar effect may be contaminating Stokes $u$. To cancel this systematic effect the above-mentioned symmetry, the reported values for the VPM transfer function were evaluated by averaging each data point with an equivalent data point taken in a configuration where the cryostat was rotated by 90$^\circ$. The resulting values for $q$ and $u$ are shown as a function of grid-mirror separation in Figure~\ref{fig:igram}A.

\subsection{Circuit Model Parameter Extraction}

We have applied the circuit model to data taken using the Hertz VPM instrument with a single VPM having wires nearly parallel to the plane of incidence. In this case, it is necessary to average over all relevant frequencies: 
\begin{equation}
\overline{D}^\prime=\int_{0}^{\infty} \overline{J}^\prime(\nu)^\dagger\overline{D}(\nu)\,\overline{J}^\prime(\nu)\psi_n(\nu) d\nu,
\end{equation}
where $\psi_n(\nu)$ is the normalized instrument bandpass (approximately $\nu_0=353\, \mu m, \,\Delta\nu/\nu=0.10$). 

The $q$ and $u$ data are fit to the VPM model using the downhill simplex method\cite{Lagarias98}.  Ten fit parameters were varied in this minimization: the angle of incidence, an offset in the grid-mirror separation, a global polarization efficiency, a HWP phase offset, $X_{LC}$, $X_L$, $R_L$, $X_C$, $X_{CC}$, and the bandwidth stretch ($\kappa$). The bandwidth stretch takes into account deviation from parallelism between the grid and mirror in the VPM.  Such an effect would cause the bandpass to appear slightly wider than that defined by the bandpass filter. 

The best fit parameters for our circuit model are given in Table 1. Where applicable, these values are compared with the values estimated from the system setup and the derived circuit values for the grid ($a=12.5\,\mu$m and $g=67.5\,\mu$m.) The resulting models for $q$ and $u$ are plotted along with the data in Figure 6A. The reduced $\chi^2$ value for the joint $(q,u)$ fit is 2.1. Upon removal of five of the $(q,u)$ pairs where either $q$ or $u$ has a significant deviation from the model, the $\chi^2$ is reduced to 1.2. From this we conclude that the circuit model captures the basic features of the polarization transfer function, however, the statistical errors are underestimated in a subset of the measurements. Such an effect can occur as a result of time-variable instabilities in the system that are insufficiently mitigated by the techniques described earlier in this section. Based upon the observed performance of the system a phase drift in the chopper wheel for a subset of the data frames is considered the most probable cause.


The resistance in the inductive mode, $R_L$, is higher than expected from the formulation given above. The expected value is based on the conductivity of gold, the coatings for the wires. This is an indication that the gold coatings are compromised; the fitted resistance is consistent with the base metal (tungsten, $\rho=1.8\times 10^{7}\,\Omega \cdot m$\cite{Pozar}).
The fitted value of $X_L$ is within 30\% of that derived from the numerical simulation described above.  We have plotted this value shown in the middle panel of Figure~\ref{fig:loss2}.  The model presented here depends on the ability to identify a unit cell for the grid; however, variations in wire spacing violate the symmetry used to calculate the properties of the grid using this technique.  Examination of the grid using a microscope confirmed this concern.


{In Figure~\ref{fig:igram}B, the features in Stokes $q$ at grid-mirror separations of approximately 200 and 400 $\mu$m are due to trapped resonances for the inductive mode. We note in passing that these features can be significantly reduced by modifying the VPM configuration. In place of the mirror, one could incorporate a second grid with wires oriented perpendicular to the first followed by beam dump. Ideally, the brightness temperature of the beam dump should match that of the source over the beam on the sky. This architectural change would reduce common-mode systematic effects between orthogonally-polarized detectors.

\begin{table}[htbp]
   \centering
   \begin{tabular}{lccccc} 
      \hline
            Parameter    & Estimate  & Fit &  Units\\ \hline
      Angle of Incidence  & 	20 & 22.83 & [degrees]	  \\
      Grid-mirror Offset  & 	0	& -2.01 & [$\mu$m]\\
      Polarization Efficiency  & 	100	& 98.4 & [\%]\\
      HWP Phase Offset & 	0	& 0.0125 &[radians] \\
      $\kappa$ 	&	 1				&1.08 & [ - ]		\\ 
      $R_L$  	&$9.9\times10^{-4\dagger}$	&	$3.4\times10^{-3}$	& [ - ] \\
       $X_L$  	&  0.031$^\dagger$		& 	0.040  & [ - ]\\
     $X_{LC}$&  -0.26$^*$	&  	-0.27 & [ - ] 	 \\
      $X_C$ 		& -		&  -0.43	& [ - ] \\
      $X_{CC}$ 	 	&-		&	-265	 & [ - ]\\ \hline
      
        \hline
   \end{tabular}
   \caption{Fit results for the 350 $\mu$m measurements. ($^\dagger${From \S 4}, 
$^*${From Marcuvitz})}

   \label{tab:parms}
\end{table}

\begin{figure}[htpb]
	\centering
	\includegraphics[width=4.5in]{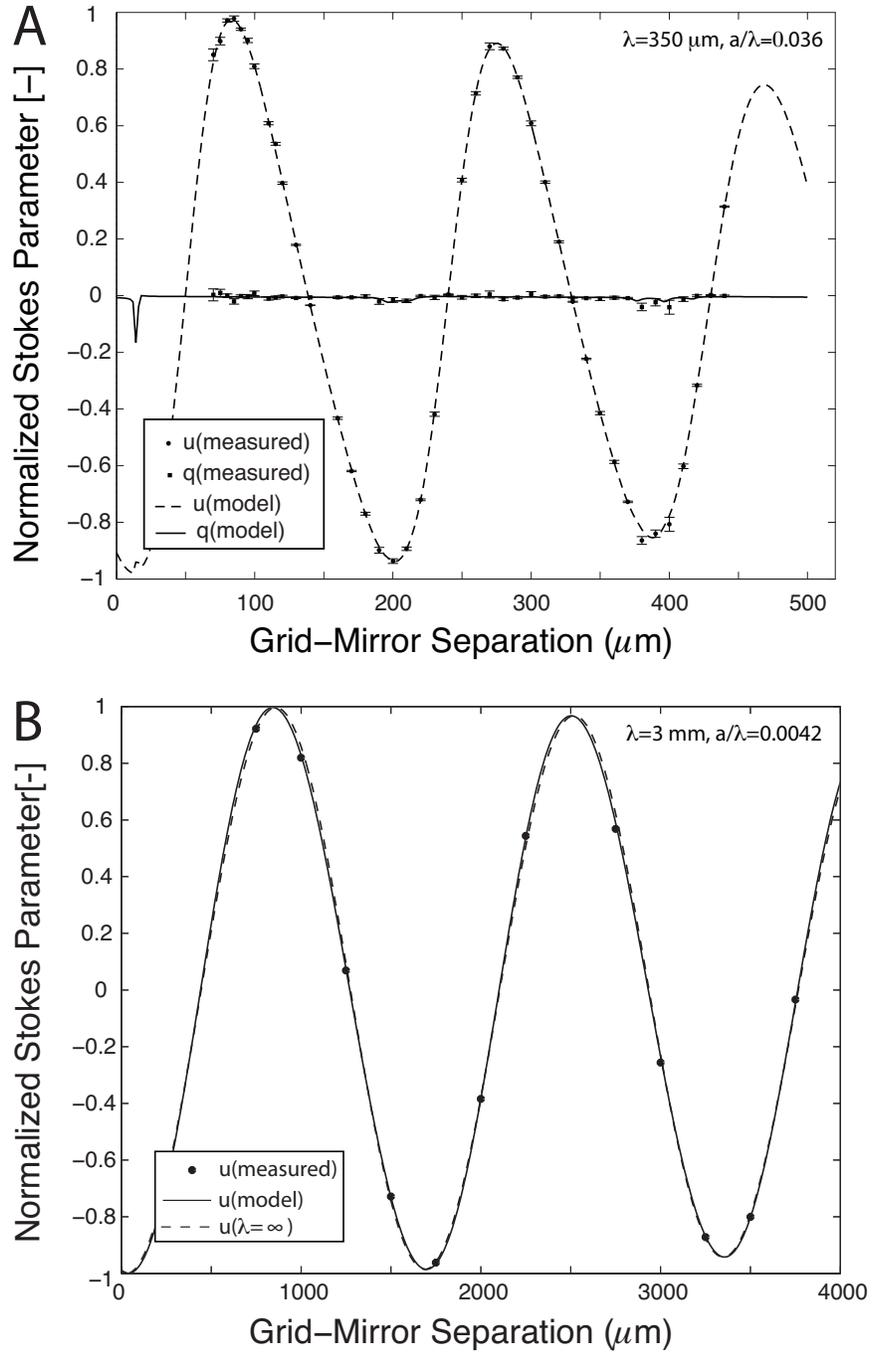}
	\caption{(A) The $q$ and $u$ values were measured as a function of grid-mirror separation at 350 $\mu$m.  (B) Measurements of the VPM at 3 mm are also shown.  The transmission line model is in close agreement with the infinite wavelength approximation at this frequency; however, the difference between the two is measurable.}
	\label{fig:igram}
\end{figure}


\subsection{VPM Modulation Efficiency}

The modulation efficiency of the Hertz polarimeter was experimentally observed to be $\sim 95\%$. This measurement was made with the source grid in place but in the absence of the VPM polarizer.  This value is consistent with previous measurements of the efficiency of the Hertz polarimeter\cite{Dowell98, Krejny08}.  For the data reported in Table~\ref{tab:parms} and Figure~\ref{fig:igram}, the reported efficiency, 98.4\%, is that of the VPM alone.     The lower modulation efficiency previously reported, 85\% \cite{Krejny08}, was found to be reproducible and induced by beam misalignments at the location of the VPM.  The estimate reported here for the modulation efficiency of the VPM is a lower limit and could arise from one of several effects. First, the circuit model assumes uniform wire spacing and size. As such, the model represents the average circuit parameters over the illuminated part of the grid and does not account for variations in parasitic coupling between the wires that are important at high frequencies \cite{Costley77, Beunen81}. Treatment of this effect would require modifications to the simple circuit model considered here.  Alternatively, and more likely, the non-unity efficiency factor could be an artifact of the system such as a residual misalignment of the beam on the VPM or residual uncertainty in the Hertz efficiency calibration.





\section{3 mm VPM Tests}
We have also measured the VPM at a wavelength of 3 mm ($g/\lambda=0.022$.) We measured the response from 75-110 GHz; however, we used the 91-100 GHz portion of the measured spectrum to synthesize a 10\% bandwidth to match the 350 $\mu$m measurement.
The setup for this test is shown in Figure~\ref{fig:wband}.  An HP 8510 vector network analyzer was used to measure the polarization state of the radiation as it is processed by the VPM.  Vertically-polarized radiation  
is emitted from port 1 and launched from waveguide to free space via a feedhorn. It is then reflected by a polarizing grid having wires oriented in the vertical direction. This grid helped to further define the polarization state of the radiation and limits the formation of trapped reflection resonances in the system. The unused polarization is terminated by a free-space beam dump. An ellipsoidal mirror is used to form a beam waist on the VPM. This ensures that the phase front of the radiation is approximately flat at the polarization modulation point. Just as for the 350 $\mu$m setup, the VPM wires were oriented such that their projected angle is 45$^\circ$ with respect to the incoming polarization state. A second, identical ellipsoidal mirror refocuses the beam on a second feedhorn that is identical to the first. A folding flat is used in this final free-space path to symmetrize the optics. The orthomode transducer (OMT) \cite{Wollack02,Wollack03} ports corresponding to the unused polarizations are terminated by precision waveguide loads. This mitigates the influence of polarization mode conversion that can result in trapped resonances in the beam waveguide structure. 

As the grid-mirror separation is varied, the VPM modulates the incoming (vertical) signal that we define as Stokes $U$ in order to maintain consistency with the 350 $\mu$m data. The modulated power is detected by port 2 which is sensitive to vertically-polarized radiation. The 800-point spectrum of $S_{12}$ over the entirety of W-band (75 to 115 GHz) is recorded at every grid-mirror separation.
Once this is finished, a 90$^\circ$ twist is added to the waveguide in port 2 to switch the port's sensitivity from vertical to horizontal polarization. The grid-mirror separation of the VPM is then adjusted in an identical way as previously described. 

\begin{figure}[htpb]
	\centering
	\includegraphics[width=4in]{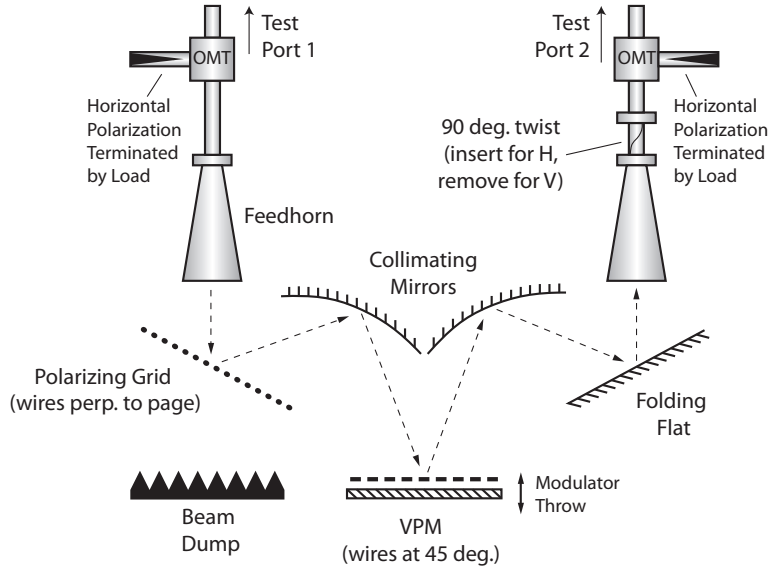}
	\caption{The setup for the 3 mm  VPM transmission test is shown. Radiation is emitted from port 1 of the HP 8510 with a vertical polarization. The radiation is then reflected off of the grid and collimated by an ellipsoidal mirror. At this point, the radiation is reflected off of the VPM. An identical ellipsoidal mirror follows, and a folding flat directs the beam into a second feed that is attached to port 2 of the HP 8510.  A 90$^\circ$ twist is added to the waveguide to change the sensitivity from the vertically-polarized state to the horizontally-polarized state. }
	\label{fig:wband}
\end{figure}

\subsection{Data Analaysis}

For comparison with the 350 $\mu$m data, we wish to determine the response of the VPM over a 10\% bandwidth fractional bandwidth.  To do this, we calculate the integrated signals $H(z)$ and $V(z)$:

\begin{align}
H(d)&\equiv\sum_{\Delta\nu} |S^H_{12}(d,\nu)|^2\\
V(d)&\equiv\sum_{\Delta\nu} |S^V_{12}(d,\nu)|^2.
\end{align}
Here, $S^H_{12}(d,\nu)$ and $S^V_{12}(d,\nu)$ are the relevant scattering parameters for the case of horizontal and vertical sensitivity for port 2, respectively.

The next step is to calculate the relative gain of the two channels. This can be done by noting that the single-frequency 
 response for the two quantities are,
 \begin{align}
 H(d)&=\frac{1}{2}[I_H+U_H\cos{\phi(d)}+V_H\sin{\phi(d)}]\\
V(d)&=\frac{1}{2}[I_V-U_V\cos{\phi(d)}-V_V\sin{\phi(d)}].
\end{align}
Here, $I_{H/V}$, $U_{H/V}$, and $V_{H/V}$ are Stokes parameters. 

For each of $H(d)$ and $V(d)$, we perform a linear least-squares fit.
Since the single-frequency will not fit our broadband response precisely, we do not use this to directly calculate Stokes parameters. Rather, we calculate a relative gain based
on the unmodulated component (Stokes I) from each fit,
\begin{equation}
f=\frac{I_H}{I_V}.
\end{equation}
From this, Stokes $u$ can be calculated:
\begin{equation}
u(d)=\frac{H(d)-fV(d)}{H(d)+fV(d)}.
\end{equation}
The results of this analysis are shown in Figure~\ref{fig:igram}B. In addition, we have plotted a VPM model using the parameters found from the 350 $\mu$m fit, appropriately scaled by wavelength.  We found that to fit the data required an incident angle of 19$^\circ$ and an offset in the grid-mirror separation of -30 $\mu$m.  No other parameters in the model have been adjusted. For comparison, we have plotted the normalized Stokes parameter,  $u$, in the infinite-wavelength limit.  In this limit, the impedance contrast is high for the inductive mode and the dependence on the circuit parameters is minimal.
  
\section{Discussion}
{Figure~\ref{fig:multfreq} summarizes the VPM transfer function for a single frequency for each of the models in this work as well as the geometric limit.  To emphasize the differences in shape of the curves, for each case we have shifted the curve to align the first minimum to lie at zero delay. For $\lambda\gg a$, the grid circuit is firmly in the single-mode limit. In this case, for plane wave illumination, Equation~\ref{eq:geom} is a good approximation for the VPM phase delay. As the wire diameter becomes a finite fraction of a wavelength, the impedance contrast of the inductive circuit is reduced. The polarization response remains a sinusoidal function of the phase delay; however, the VPM reflection phase is dependent upon the details of the grid geometry and impedance contrast.  

\section{Summary}
We have applied a transmission line model to the VPM used for the Hertz/VPM submillimeter polarimeter and have obtained estimates for key circuit parameters using numerical simulations. The simulations represent an improvement over analytical approximations in estimating the inductance of wire grid polarizers having $2\pi a\sim g$, such as those used in this work. 

At 350 $\mu$m the VPM grids have wires that are $\sim$0.07$\lambda$ in diameter, corresponding to a free-space phase shift across the width of a wire that exceeds 10$^\circ$. We have shown that in this regime, it is not sufficient to identify the VPM phase delay with that derived from the path difference induced by the grid-mirror separation.  The VPM transfer function is modified from the infinite wavelength case, both in shape and in position. This is due to a combination of the reactances imposed by the thickness and spacing of the wires and the interaction of these reactance with the variable position of the mirror. At 3 mm, where the impedance contrast for the inductive mode is higher, differences from the geometric phase delay (Eq.~\ref{eq:geom}) are reduced, but evident.

\begin{figure}[htpb]
	\centering
	\includegraphics[width=5in]{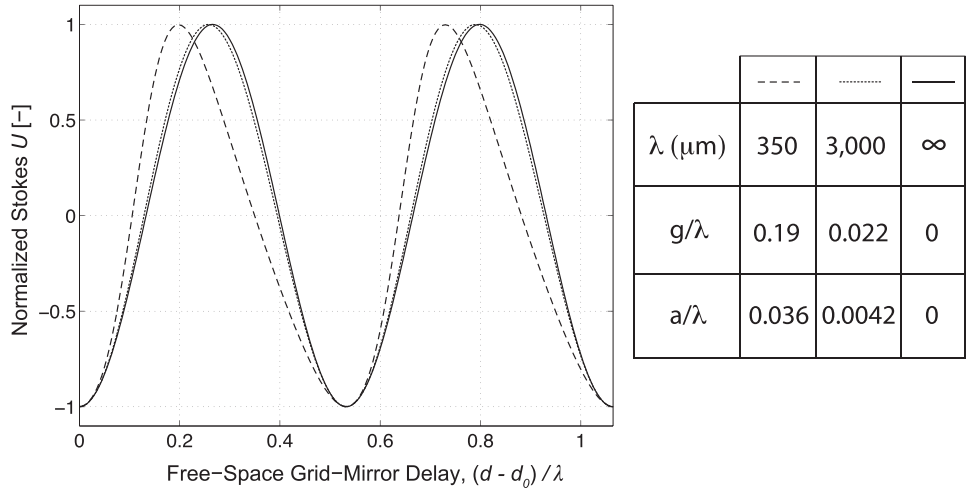}
	\caption{The modeled VPM reflection phase delay is shown for single-frequencies at 350 and 3000 $\mu$m for the grid models obtained above. In the limit $g/\lambda\ll 1$, a sinusoidal form for Stokes $U$ is observed. As this condition is relaxed, the VPM reflection phase delay differs from the free-space grid-mirror delay. The legend on the right shows the parameters corresponding to each of the curves. }
	\label{fig:multfreq}
\end{figure}

\section*{Acknowledgements} This work was funded by a NASA ROSES/APRA award. M. Krejny was supported
by Graduate Student Researchers Program grant NNG05-GL31H. H. Hui was supported by an Undergraduate Student Research Program grant
at Goddard Space Flight Center. We would like to thank G. Voellmer for his work on design and construction of the VPMs. We would also like to thanks R.F. Loewenstein, C. Walker, C. Kulesa, C.Y.  Drouet d'Aubigny, and D. Golish for their work on Hertz/VPM. We thank Roger Hildebrand for the use of the Hertz cryostat and Bob Pernic for critical cryostat repairs during the laboratory testing. We also thank the anonymous reviewers whose comments have helped to improve the paper. 
 
\clearpage
\bibliographystyle{osajnl}

\end{document}